\documentclass[prd,aps,amsfonts,eqsecnum,superscriptaddress,nofootinbib,longbibliography,notitlepage]{revtex4-1}

\usepackage{graphicx}
\usepackage{xcolor}
\usepackage{rotating,comment}
\usepackage{amsmath,amssymb,graphics,amsthm,isomath}

\usepackage[colorlinks=true, urlcolor=violet, linkcolor=blue, citecolor=red, hyperindex=true, linktocpage=true]{hyperref}
\usepackage[capitalise,compress]{cleveref}

\makeatletter
\renewcommand{\p@subsection}{}
\renewcommand{\p@subsubsection}{}
\makeatother

\usepackage{xcolor}
\usepackage{mathtools}

\usepackage{dsfont}

\begin{document}

\title{Renormalization group in quantum critical theories with Harris-marginal disorder}

\author{Koushik Ganesan}
\email{koushik.ganesan@colorado.edu}
\affiliation{Department of Physics and Center for Theory of Quantum Matter, University of Colorado, Boulder CO 80309, USA}

\author{Andrew Lucas}
\email{andrew.j.lucas@colorado.edu}
\affiliation{Department of Physics and Center for Theory of Quantum Matter, University of Colorado, Boulder CO 80309, USA}

\author{Leo Radzihovsky}
\email{radzihov@colorado.edu}
\affiliation{Department of Physics and Center for Theory of Quantum Matter, University of Colorado, Boulder CO 80309, USA}

\begin{abstract}
We develop a renormalization group for weak Harris-marginal disorder in otherwise strongly interacting quantum critical theories, focusing on systems which have emergent conformal invariance.  Using conformal perturbation theory, we argue that previously proposed random lines of fixed points with Lifshitz scaling in fact flow towards other universal fixed points, and this flow is captured by a ``one-loop" analysis. Our approach appears best controlled in theories with only a few operators with low scaling dimension.  In this regime, we compare our predictions for the flow of disorder to holographic models, and find complete agreement.
\end{abstract}

\date{\today}

\maketitle
\tableofcontents

\section{Introduction}

\subsection{Disorder and strong interactions}
A longstanding challenge in understanding the landscape of phases of quantum matter has been a careful analysis of the interplay between disorder and strong interactions.  While in some contexts, disorder simply tends to prevent phase ordering or symmetry breaking \cite{Larkin1970,ImryMa1975}, it is also possible for relevant disorder to lead to entirely new phases of matter, such as glasses \cite{FisherHuseSpinGlass1988}, Anderson \cite{AndersonLocalization} or many-body localization \cite{FleishmanAnderson1980,Basko_2006,GornyMirlinPolyakov2005,OH2007} or ``infinite randomness" fixed points \cite{RTFIMfisher1992,RAFMfisher1994}.

It was first proposed in \cite{Hartnoll:2014cua} that, at least in the context of a specific holographic model, yet another possibility might arise:  a strongly correlated quantum critical system, (weakly) perturbed by Harris-marginal disorder \cite{Harris_1974}, would flow to a ``Lifshitz fixed point" -- a non-relativistic scaling theory where the dynamical critical exponent $z$ characterizing the scaling asymmetry between time and space could continuously depend on disorder strength $D$: \begin{equation}
    z = 1+ cD + \mathrm{O}(D^2)
\end{equation}where $c>0$ is a non-universal prefactor.  The possible existence of this new line of fixed points led to many further investigations, which also found this same physics in holographic models \cite{Hartnoll:2015faa,Hartnoll:2015rza,OKeeffe:2015qma,Garcia-Garcia:2015crx} and even from generic field theoretic arguments \cite{Aharony:2018mjm}.

However, as two of us recently pointed out \cite{Ganesan:2020wzm}, and as we will sketch out more quantitatively in the following subsection, the holographic argument of \cite{Hartnoll:2014cua} appears inconsistent at sufficiently low energy scales. Instead, we argued that the line of Lifshitz ``fixed points" found previously is in fact a transient effect: the disorder (and thus exponent $z$) do in fact flow to particular values.  The purpose of this paper is to revisit our argument from a field theoretic perspective, and to confirm and generalize our original conclusions: in general, marginal disorder will flow either to 0 or to large values, and there is no line of (marginally) stable Lifshitz fixed points.

\subsection{Summary of results}
We start with a ``UV" replicated action that contains all length scales $L\gtrsim \Lambda^{-1}$.   Following the Wilsonian paradigm, we wish to integrate out short length scales and obtain an effective action directly for the physics at scale $L\gg \Lambda^{-1}$.   However, we cannot directly proceed using Feynman diagrammatics, since we are dealing with a strongly interacting theory which might, in fact, not have a local Lagrangian.  (An example of such a theory would be the O(2) Wilson-Fisher theory in $2+1$ dimensions.)

Instead, we will make simple arguments, based both on careful scaling analyses together with conformal perturbation theory (see e.g. \cite{Amoretti:2017aze}), to develop a simple ``one-loop" prescription for analyzing the RG flow of the disorder strength $D$, along with the dynamical critical exponent $z$. We first show that in certain CFTs with a conserved U(1) charge in $2+1$ dimension, the disorder is marginally relevant. This is a particularly interesting case to study, since due to rather generic non-renormalization theorems for the scaling dimension of the charge density \cite{PhysRevB.46.2655,Sachdev_1994}, the density operator must be Harris-marginal even in a strongly coupled system. In this case, at energy scale $E$, we will show that
\begin{equation}
    D(E) = \dfrac{D(\Lambda)}{\displaystyle 1 - D(\Lambda) \frac{|C_{\mathcal{JJ}T}|}{C_{TT}}\log \frac{\Lambda}{E}} \label{eq:introD1}
\end{equation}
We then compare these results to a minimal holographic model, and find agreement. 

As a second example, we study CFTs coupled to Harris-marginal (Lorentz-) scalar disorder.  Here, we find that the scalar disorder is \emph{marginally irrelevant}: the effective disorder coupling strength is given by 
\begin{equation}
    D(E) = \frac{D(\Lambda)}{\displaystyle1+D(\Lambda) \frac{|C_{\mathcal{OO}T}|}{C_{TT}}\log \frac{\Lambda}{E}}. \label{eq:introD2}
\end{equation}
In (\ref{eq:introD1}) and (\ref{eq:introD2}), $C_{\mathcal{JJ}T}$, $C_{\mathcal{OO}T}$ and $C_{TT}$ are the standard normalizations of three and two point functions in position space whose ratio is constrained by Ward identities \cite{OSBORN1994311,Aharony:2018mjm}; details will be provided later as appropriate.

A little more precisely, let $\mathcal{J}$ be the timelike component of a conserved spin-1 current, and $\mathcal{O}$ a scalar operator.  These are the operators which we source with Harris-marginal disorder.  The assumptions which then go in to our simple calculation are that the dominant contribution to the $\mathcal{J}_t\mathcal{J}_t$ and $\mathcal{OO}$ operator product expansion is the stress tensor $T^{\mu\nu}$. While in a theory such as the 2+1D Wilson-Fisher fixed point, there is no reason for this to hold, this property \emph{is} generic in CFTs described by simple holographic models, such as scalars minimally coupled to Einstein gravity. Therefore, our minimalist conformal perturbative approach is sufficient to compare with the holographic computation of \cite{Ganesan:2020wzm}, where we find complete agreement.  We will present an analogous holographic calculation with charge disorder, and again find agreement with our field theoretical arguments.

The rest of this paper is organized as follows: In section \ref{sec:charge} we introduce our charge disorder model and obtain the replicated action. Section \ref{sec:previ} reviews the earlier argument of \cite{Aharony:2018mjm} for the emergence of Lifshitz fixed points in generic systems with Harris-marginal disorder.  Section \ref{sec:cpt} and \ref{sec:holo_vec} contains our ``one-loop" perturbative and holographic computations for the flow of $D$ in the vector case respectively. In section \ref{sec:scalar}, we perform a similar field theoretic calculcation for the scalar case and compare the results to \cite{Ganesan:2020wzm}. Section \ref{sec:conclusions} discusses possible applications and extensions of our results, along with a comparison to some other work.

\section{Charge disorder in two spatial dimensions: field theory}\label{sec:charge}
We begin by discussing charge disorder in two spatial dimensions. This ``dirty boson" problem in condensed matter physics has been extensively explored recently in  \cite{Kim_1994,Goldman_2020}.  Our calculation is somewhat different: as stated above, our approach is valid even in a strongly coupled field theory (without an $\epsilon$ or large $N$ expansion); however, we neglect contributions to the operator product expansion which might be important in an $N=1$ theory.  
\subsection{Model details}
We consider a quantum critical theory, described by a conformal field theory (CFT) in $d$ spatial dimensions.  This means that the dynamical critical exponent $z=1$, where $z$ is defined by the relative scaling of time and space: $[t]=z[x]$.  We also assume hyperscaling: $\theta=0$, hence the free energy $F(T)\sim T^{1+d/z}\sim T^{1+d}$.   In what follows, we focus on zero temperature physics, and work in Euclidean time.  

Suppose that this CFT has a U(1) conserved charge and therefore an spin-1 conserved current $\mathcal{J}$ of dimension 
\begin{equation}
   [ \mathcal{J}] = \Delta = d
\end{equation}
We will relax this assumption in section \ref{sec:scalar}. Then for such an a vector operator,
\begin{equation}\label{jjcorr}
\langle \mathcal{J}_{\mu}(x)\mathcal{J}_{\nu}(0)\rangle_{\mathrm{CFT}} = \frac{C_{\mathcal{JJ}}(\delta_{\mu\nu}-\frac{2x_\mu x_\nu}{x^2})}{x^{2d}}
\end{equation}
Now consider perturbing the CFT Lagrangian density by a space-dependent but time-independent coupling $h(x)$, physically corresponding to a quenched random potential, due to e.g., sample impurities: 
\begin{equation}\label{eq:per_act}
    \mathcal{L} \rightarrow \mathcal{L}+ \mu(x)\mathcal{J}_t(x,t)
\end{equation}
where $h^t = \mu$. Note that $[\mu]=1$. We assume that $\mu(x)$ is drawn from a Gaussian distribution with zero mean and covariance matrix 
\begin{equation}\label{eq:harris_dis}
    \overline{\mu(x)\mu(x')} = 2\pi D \times G(\Lambda|x-x'|)
\end{equation}
$D$ corresponds to the strength of the random potential, and is dimensionless as the disorder is Harris-marginal. We include an extra factor of $2\pi$ to account for the field normalization that is simpler in holographic models. The UV cutoff is given by $\Lambda$, and $G$ is a smooth function such as $G(r)\sim \mathrm{e}^{-r}$.  We normalize the dimensionful $G$ (note that $[G]=d$) so that \begin{equation}
    1 = \int\mathrm{d}^dx G(\Lambda|x|).
\end{equation}
On long wavelengths, therefore, $G(\Lambda|x|) \approx \delta^{(d)}(x)$. 

The inverse length scale $\Lambda$ corresponds to a UV cutoff \emph{on disorder} (namely, the inverse of its correlation length).  Within the $z=1$ fixed point theory, of course, $\Lambda$ is also the UV energy scale below which the disorder may qualitatively change the physics. We must emphasize that there is a separate energy scale $\Lambda_{\mathrm{CFT}}\gg \Lambda$ above which the continuum field theory description no longer holds (at least if we assume that the true microscopic physics takes place in a lattice model).  We will not be interested in adjusting $\Lambda_{\mathrm{CFT}}$ because, by assumption, the CFT is an IR fixed point when $D=0$; hence, for the discussion that follows, we will treat the ``UV" theory to be the CFT.  Our goal is to determine the IR fixed point once the CFT is perturbed by Harris-marginal disorder.

We will analyze this theory using the replica method (although the ``peculiar" $n\rightarrow 0$ limit does not appear important for the calculation).  For pedagogical purposes, let us illustrate our procedure by assuming some ``Lagrangian" exists for this theory (though, ultimately, this assumption does not factor in to our calculation).  Let $a,b$ denote replica indices. Upon disorder averaging (\ref{eq:per_act}) using (\ref{eq:harris_dis}), we obtain the effective action 
\begin{equation}\label{eq:seff}
    S_{\mathrm{eff}} = \sum_{a=1}^n \int \mathrm{d}^dx \mathrm{d}t\; \mathcal{L}_a- \frac{2\pi D}{2} \sum_{a,b=1}^n \int \mathrm{d}^dx_1\mathrm{d}^dx_2 \mathrm{d}t_1\mathrm{d}t_2 \;G(\Lambda|x_1-x_2|) \mathcal{J}^a_t(x_1,t_1)\mathcal{J}^b_t(x_2,t_2) = \sum_{a=1}^n S_{\mathrm{CFT},a} + S_{\mathrm{dis}}
\end{equation}
where $\mathcal{L}_a$ denotes the CFT Lagrangian density for replica $a$, and only depends on the $a$-copy of the fields. 

Since we require our disorder to be Harris-marginal ($D$ is dimensionless), we set $d=2$ for the reminder of the analysis on charge disorder.

\subsection{Emergence of Lifshitz scaling}\label{sec:previ}
Let us now review the argument of \cite{Aharony:2018mjm} for the emergence of Lifshitz scaling in the presence of disorder. Even though the original argument was presented for scalar disorder, an analogous approach works for vector disorder that we consider in this section. We focus on the second term in (\ref{eq:seff}). When we bring two operators very close, it is reasonable to try and use the operator product expansion (OPE) to simplify this expression. One universal contribution to this OPE arises due to the stress tensor:
\begin{equation}\label{eq:JOPE}
    \mathcal{J}^a_t(x_1,t_1)\mathcal{J}^b_t(x_1,t_2) = \frac{\delta_{ab}C_{\mathcal{JJ}T}}{C_{TT}} \frac{T_{tt}(x_1,t_1)}{|t_{12}|}  + \cdots 
\end{equation}
where $t_{12} = t_1-t_2$. Note that the form of this OPE is fixed by conformal invariance.  Now, suppose that in the textbook spirit of Wilson's RG, we wish to reduce our UV cutoff from $\Lambda$ to $\Lambda/b$, with $b>1$ a small parameter. Integrating over $t_{12}$ we find 
\begin{align}
    S_{\mathrm{eff}} &\approx \cdots - \frac{2\pi D}{2} \sum_{a=1}^n\int \mathrm{d}^2x \mathrm{d}t \; T^a_{tt}(x,t) \times \int^{\frac{b}{\Lambda}}_{\frac{1}{\Lambda}} \mathrm{d}t^\prime\;\frac{C_{\mathcal{JJ}T}}{C_{TT}} \frac{1}{|t'|}. 
\end{align}
The logarithmically IR-divergent integral above leads to non-trivial critical exponents.  In particular, observe that
\begin{equation}
    S_{\mathrm{eff}} = \cdots - 2\pi D\frac{C_{\mathcal{JJ}T}}{C_{TT}} \log b \int \mathrm{d}t H(t), \label{eq:sefflifshitz}
\end{equation}
where $H(t)$ is the total Hamiltonian (i.e. energy) in the system.\footnote{Note that since $t^\prime$ can be positive or negative, the integral over $t^\prime$ leads to a factor of $2\log \Lambda$.} This divergence can be effectively cancelled by rescaling time coordinate as $t \xrightarrow[]{} tZ_t$, since the variation of the action under this transformation takes the form $\delta S = -(Z_t-1)\int \mathrm{d}t H$. For us, we see that:
\begin{equation}\label{eq:Zt}
    Z_t = 1-2\pi D\frac{C_{\mathcal{JJ}T}}{C_{TT}}\log b
\end{equation}
This suggests that the theory has Lifshitz invariance, since
\begin{equation}
    \frac{\mathrm{d} Z_t}{\mathrm{d}\log b} =  - 2\pi D\frac{C_{\mathcal{JJ}T}}{C_{TT}} = z-1. \label{eq:lifshitz}
\end{equation}
So we expect that time scales a little differently from space: $z=1$ is broken by an order $D$ correction.

Note that we do not need to consider other components of the stress tensor in the OPE.  The nonlocality in (\ref{eq:seff}) is only in the time direction. After performing the integral over $t_{12}$, we find that the only divergence arises in the energy channel.




\subsection{Flow of disorder}\label{sec:cpt}
The main point of this paper is to show that the argument above is incomplete: $D$ will generically flow under RG.  As such, there will in general be no line of fixed points with disorder-dependent dynamical critical exponent.

In order to find the flow equation for $D$, we need to keep in mind that the coupling constants in (\ref{eq:seff}) are the ``bare" couplings.  They should be re-scaled as follows:
\begin{subequations}\begin{align}
    \mathcal{J}^{\text{bare}}_t &= \frac{\mathcal{J}^R_t}{\sqrt{Z_{\mathcal{JJ}}}}\\
    t^{\text{bare}} &= t^RZ_t\\
    D^{\text{bare}} &= D^R\frac{Z_{\mathcal{JJ}}}{Z_t^2} \label{eq:J_D_flow}
\end{align}\end{subequations}
where $Z_i = 1+\delta_i$, with $\delta_i$ corresponding to couplings of counter terms.
We already calculated $Z_t$ in (\ref{eq:Zt}) and we will determine $Z_{\mathcal{JJ}}$ shortly. 

Clearly, we cannot implement the RG sketched above using a conventional diagrammatic expansion. We may not even have a controlled Lagrangian description of the CFT, as is the case in even a simplest non-trivial example ($\mathrm{O}(2)$ Wilson-Fisher fixed point in $2+1$ dimensions). 
Still, in the spirit of the canonical Wilsonian RG, to leading order, (\ref{eq:lifshitz}) generalizes to give us 
\begin{equation}
    -\frac{\mathrm{d}\log t}{\mathrm{d}\log b} = 1 - 2\pi D(E)\frac{C_{\mathcal{JJ}T}}{C_{TT}} + \cdots \label{eq:dlogt}
\end{equation}
where $\cdots$ represents higher order terms in $D$, which are beyond the scope of this work.
We claim that
\begin{equation}\label{eq:ZJJ}
    Z_{\mathcal{JJ}} = 1
\end{equation} since we know from non-renormalization theorems \cite{Sachdev_1994, PhysRevB.46.2655} that a conserved current does not gain an anomalous dimension unless the conservation law is broken. We should emphasize that there are certain holographic models that do support anomalous dimensions for $\mathcal{J}$ \cite{PhysRevB.91.155126,Gouteraux:2013oca,Karch:2014mba} but these models typically involve introducing other fields (such as a dilaton) coupled to the gauge field or breaking of the gauge symmetry.

Now combining (\ref{eq:J_D_flow}), (\ref{eq:Zt}), and (\ref{eq:ZJJ}) we find,
\begin{equation}
    D^{\text{bare}} = D\frac{1}{Z_t^2}
\end{equation}
Hence,
\begin{equation}
    b \frac{\partial D}{\partial b} = -\frac{4D^2\pi C_{\mathcal{JJ}T}}{C_{TT}}
\end{equation}
At energy scale $E$, which is defined by $b=\Lambda/E$, we conclude that the effective disorder strength is given by
\begin{equation}\label{dflow}
    D(E) = \dfrac{D(\Lambda)}{\displaystyle 1 +2\pi D(\Lambda) \frac{2C_{\mathcal{JJ}T}}{C_{TT}}\log \frac{\Lambda}{E}}
\end{equation}
We can also see what happens to the flow of the dynamical critical exponent $z$.  Upon integrating (\ref{dflow}) using (\ref{eq:Dflow}), we find that \begin{equation}
   -\frac{\mathrm{d}\log t}{\mathrm{d}\log b} = 1 - \dfrac{\displaystyle 2\pi D(\Lambda)\frac{C_{\mathcal{JJ}T}}{C_{TT}}}{\displaystyle 1 + 2\pi D(\Lambda) \frac{2C_{\mathcal{JJ}T}}{C_{TT}}\log b}
\end{equation}
which means that 
\begin{equation}
    -\log t = \log b - \frac{1}{2}\log \left( 1 + 2\pi D(\Lambda) \frac{2C_{\mathcal{JJ}T}}{C_{TT}}\log b\right)
\end{equation}
The effective dynamical critical exponent at energy scale $E<\Lambda$ is given by  
\begin{equation}\label{eq5.26}
    z_{\mathrm{eff}}(E) = -\frac{\log t}{\log b} = 1 - \dfrac{\displaystyle \log \left( 1 +2\pi D(\Lambda) \frac{2C_{\mathcal{JJ}T}}{C_{TT}}\log \frac{\Lambda}{E}\right)}{\displaystyle 2\log \frac{\Lambda}{E}}
\end{equation}

Now using the results for the three-point functions of a conserved current found in \cite{Aharony:2018mjm,OSBORN1994311,PhysRevD.60.026004,FREEDMAN199996}, we find that
\begin{align}\label{eq:value_CJJT_CT}
    \frac{C_{\mathcal{JJ}T}}{C_{TT}} &= -\frac{3}{64\pi}.
\end{align}
which gives,
\begin{equation}\label{eq:z_vector}
    z(E) = 1 - \dfrac{\displaystyle \log \left( 1 -\frac{3D(\Lambda)}{16}\log \frac{\Lambda}{E}\right)}{\displaystyle 2\log \frac{\Lambda}{E}}
\end{equation}
We learn from (\ref{dflow}) that the disorder becomes marginally relevant, because $C_{\mathcal{JJ}T}<0$.   There is simultaneously a blow up in the dynamical critical exponent at an IR scale \begin{equation}
    \Lambda_{\textrm{IR}}\sim \Lambda \mathrm{e}^{-\frac{16}{3D(\Lambda)}}.
\end{equation} 

This exponential scaling of an IR scale at which perturbation theory breaks down is reminiscent of the localization length scale in Anderson localization in two spatial dimensions for a system with time-reversal symmetry \cite{PhysRevLett.42.673}.  It is tempting to speculate that holographic systems in four bulk spacetime dimensions with charge disorder that there may also be a similar kind of ``localization" phenomenon, perhaps even associated with fragmentation of the bulk horizon \cite{Anninos:2013mfa,Horowitz:2014gva}.  We would predict that this phenomenon is already visible within the Einstein-Maxwell theory, as studied above.  However, a curious known fact is that (at least until the horizon fragments) the conductivity of such holographic models is always finite \cite{Grozdanov:2015qia,PhysRevD.93.061901}.  One possible resolution is that fragmentation does not happen, and that there is some non-perturbative resolution of the disorder in holographic models leading to a smooth horizon;  or, there may be a first order phase transition to the fragmented/localized phase where the conductivity abruptly drops to zero; or, it may be the case that localization is in a sector of the theory decoupled from charge \emph{and} heat transport (namely the ``localized" highly disordered phase remains a conductor). It would be interesting to study this problem further.

\section{Charge disorder in two spatial dimensions: holographic model}\label{sec:holo_vec}
In general, it is far from clear that the OPE channel $\mathcal{JJ}T$ is the only contribution to the RG flow of $D$. However, there is one class of CFTs for which we can assure this is the case: a CFT described by gauge-gravity duality \cite{Hartnoll:2016apf}.  In a nutshell, the holographic correspondence states that the CFT (if it is of a suitable large $N$ matrix type) is dual to a semiclassical gravity theory in asymptotically anti-de Sitter (AdS) space in one higher dimension.  The conserved current $\mathcal{J}$ is dual in the bulk to a gauge field $A$, and the stress tensor $T_{\mu\nu}$ is dual to the spacetime metric $g_{\mu\nu}$.  The holographic action controls CFT data, such as scaling dimensions and OPE coefficients.  


A classic example of such a CFT is the ABJM theory in 2+1 spacetime dimensions \cite{Aharony:2008ug}.  This theory has a conserved U(1) current, and it is holographically realized by the $\mathrm{AdS}_4$-Einstein-Maxwell theory with gravitational action:
\begin{equation}
    S_{\mathrm{Grav.}} = \int \mathrm{d}t\mathrm{d}^2x\mathrm{d}r\sqrt{-g}\bigg(R+6-\frac{1}{4}F_{ab}F^{ab}\bigg)
\end{equation}
where we have set the gravitational constant and AdS radius to 1 without loss of generality. The indices $a$,$b$ denote bulk coordinates with $r$ being the bulk radial direction. Since we want the metric to be asymptotically AdS, we have set the cosmological constant to $-3$, in units where the AdS radius has been set to 1.

\subsection{Constructing the bulk geometry}
In holography, the Harris-marginal disorder (\ref{eq:harris_dis}) corresponds to setting boundary conditions on the timelike component of the gauge field $A_t$ near the boundary of AdS.  By solving the bulk equations of motion carefully, we can hope to track the RG flow of the coupling constants, following the prescription of \cite{Ganesan:2020wzm}. In particular, we will try to directly construct a homogeneous approximation to the bulk geometry which is self-consistent.  There are no assumptions about scaling dimensions and non-renormalization theorems: instead, the consistent solutions to the gravitational equations will encode this physics.

The coupled equations of motion for the metric tensor and the gauge field are
\begin{subequations}
\begin{align}
    \nabla_aF^{ab} &= 0\\
    R_{ab}-\frac{1}{2}Rg_{ab}-3g_{ab} &= \frac{1}{2}\bigg(F^c_aF_{cb}-\frac{1}{4}g_{ab}F^2\bigg)\label{5b}
\end{align}
\end{subequations}
Since the background field theory is initially disorder free, the solution on the gravity side, at $r=0$, corresponds to an asymptotically $AdS_4$ geometry:
\begin{subequations}
\begin{align}
    \mathrm{d}s^2 (r\rightarrow 0) &= -\frac{\mathrm{d}t^2}{r^2}+\frac{\mathrm{d}r^2}{r^2}+\frac{\mathrm{d}\vec{x}^2}{r^2}\\
    \mathcal{A}_t(\vec{x},r) &= \mu(\vec{x}) + \mathcal{J}_t(\vec{x})r+..
\end{align}
\end{subequations}
where we have used standard quantization. Thus the coefficient of the leading power of $r$ ($\mu(x)$) in the near boundary expansion of $\mathcal{A}_t$ corresponds to the chemical potential and sub-leading correction ($\mathcal{J}_t(x)$) to the charge density of the dual field theory. In our calculations we work in the temporal gauge $A_t(r,x,y) = \mathcal{A}_t(r)e  ^{ik_xx+ik_yy}$, and find that no other components of the gauge field are sourced in the bulk by the equations of motion.

In general, solving these equations exactly is extraordinarily difficult and can only be done numerically.  Yet, we have already seen that we must go to \emph{exponentially small} energy scales $\Lambda_{\mathrm{IR}}$ (which will roughly mean exponentially large $r$)  in order to understand the RG flow of the Harris marginal disorder.  Hence, it is unlikely that even the best present day numerical techniques \cite{Dias:2015nua} would suffice to resolve the IR geometry.  We therefore proceed using an approximation technique from \cite{Hartnoll:2016apf,Ganesan:2020wzm}.  We assume the metric takes the form
\begin{equation}\label{eq:metric_charge_ansatz}
    \mathrm{d}s^2 = A(r)\mathrm{d}r^2+\frac{\mathrm{d}\vec{x}^2}{r^2}-B(r)\mathrm{d}t^2.
\end{equation}
Note that this homogeneity assumption is reasonable: inhomogeneity in the metric is $\mathrm{O}(D)$, and feeds back in to affect the homogeneous part of the metric only at $\mathrm{O}(D^2)$.  As long as $D$ is small therefore, we do not expect this ansatz will cause problems.  To calculate $A$ and $B$, we disorder average the stress energy tensor in (\ref{5b}) prior to solving for the metric components. After some algebra, Einstein's field equations become
\begin{subequations}\label{eq:einstein's_field}
\begin{align}
    rA\partial_r\bigg(\frac{1}{r^4A}\bigg)-\frac{\partial_rB}{r^3B} &= \frac{A}{B}\overline{(\partial_x\mathcal{A}_t)^2}\\
    6A-\frac{6}{r^2}+\frac{A}{r B}\partial_r\bigg(\frac{B}{A}\bigg) &= \frac{\overline{(\partial_r\mathcal{A}_t)^2}}{2B}\\
    \frac{8}{r^2}-12A+\frac{2B}{rA}\partial_r\bigg(\frac{A}{B}\bigg)+2\sqrt{\frac{A}{B}}\partial_r\bigg(\frac{\partial_rB}{\sqrt{AB}}\bigg) &= \frac{\overline{(\partial_r\mathcal{A}_t)^2}}{B}
\end{align}
\end{subequations}
where we have assumed the disorder averages are isotropic along the spatial directions: $\overline{(\partial_x\mathcal{A}_t)^2}=\overline{(\partial_y\mathcal{A}_t)^2}$.

To solve (\ref{eq:einstein's_field}), the first step is to evaluate the disorder averages. For this we need to solve the gauge field equation of motion for a gauge field at spatial wave number $k$ (dual to the disorder of the same wave number) given by
\begin{equation}\label{eq:holo_gauge_field_eq}
    \partial_{\mu}(\sqrt{-g}g^{\mu\nu}g^{tt}\partial_\nu \mathcal{A}_t(r,x,y)) = 0
\end{equation}
We use the following metric ansatz
\begin{subequations}\label{eq:metric_ansatz}
\begin{align}
    A(r) &= \frac{c}{r^2}\\
    B(r) &= \frac{1}{r^2(r\Lambda)^{2z_{\mathrm{eff}}(r)-2}}
\end{align}
\end{subequations}
where we assume the geometry varies sufficiently slowly in the radial direction: in particular, \begin{equation}\label{eq:holodrsmall}
    r\partial_rz_{\mathrm{eff}}(r)\ll 1.
\end{equation} 
If so, then the solution to (\ref{eq:holo_gauge_field_eq}) is
\begin{equation}\label{eq:gauge_field_eq_holo_soln}
    \mathcal{A}_t(r) \approx \sqrt{2\pi}(r\Lambda)^{\frac{1-z_{\mathrm{eff}}(r)}{2}}\mathrm{e}^{-\sqrt{c}kr}
\end{equation}
where the $\Lambda$ dependence comes in because the metric transitions to AdS close to $r\approx\frac{1}{\Lambda}$. This tells us that the dimension of the source has now become $[\mu(x)] = z_{\mathrm{eff}}$. We therefore find that the disorder averages go as
\begin{subequations}\label{eq:disorder_averages_holo}
\begin{align}
    \frac{A}{B}\overline{(\partial_x\mathcal{A}_t)^2} &= (r\Lambda)^{z_{\mathrm{eff}}(r)-1}  \cdot  c \frac{3D}{16c^2r^4}\\
    \frac{1}{2B}\overline{(\partial_r\mathcal{A}_t)^2} &= (r\Lambda)^{z_{\mathrm{eff}}(r)-1} \cdot \frac{r^2}{2} \frac{D(24+4(z_{\mathrm{eff}}(r)-1+r\log(r\Lambda)z'_{\mathrm{eff}}(r)))(3+z_{\mathrm{eff}}(r)+r\log(r\Lambda)z'_{\mathrm{eff}}(r))}{64cr^4}
\end{align}
\end{subequations}
Note that the left hand side of (\ref{eq:einstein's_field}) does not pick up any anomalous factor of $r\Lambda$ in the IR.  As a consequence, we will find that the right hand sides of (\ref{eq:einstein's_field}) become very large in the IR, since $z_{\mathrm{eff}}\ge 1$, and the geometry must qualitatively change to account for this.  Solving the set of coupled equations in (\ref{eq:einstein's_field}) at moderately small values of $r$ (i.e. not too deep in the IR) using (\ref{eq:metric_ansatz}) and (\ref{eq:disorder_averages_holo}), following \cite{Ganesan:2020wzm}, we obtain 
\begin{subequations}\label{eq:holo_z}
\begin{align}
    c &= 1+\frac{D}{16}\\
    z_{\mathrm{eff}}(r) &\sim 1-\frac{\log\bigg(1-\frac{3D}{16}\log(r\Lambda)\bigg)}{2\log(r \Lambda)}
\end{align}
\end{subequations}
with $\Lambda$  the UV cutoff. This is precisely what we expect to find for marginally relevant disorder as in (\ref{eq5.26}).

This suggests that there is an emergent IR scale where $z(r)$ blows up:
\begin{equation}
    \Lambda_* \sim \Lambda \mathrm{e}^{-\frac{a}{D}}
\end{equation}
where $a$ is an $O(1)$ constant which we estimate is close to $\frac{8}{3}$. The assumption (\ref{eq:holodrsmall}) is thus justified as long as $r<\frac{1}{\Lambda_*}$. For energy scales $E$ obeying $\Lambda_*\ll E\ll\Lambda$, we have an effective Lifshitz metric with the dynamical critical exponent locally given by 
\begin{equation}
    z = 1+\frac{3}{32}D+\cdots.
\end{equation}
Of course, this value of $z$ is not really constant, but rather slowly varying.

The blow up of (\ref{eq:holo_z}) is not surprising as the disorder has become relevant in the IR of the Lifshitz geometry. A blow up in the metric due to the presence of relevant disorder in $\mathrm{AdS}_4$ geometry was first proposed in \cite{Adams:2012yi}.  Later,  \cite{OKeeffe:2015qma} explained how to remove these divergences and arrive at a well behaved solution with Lifshitz scaling.  Amusingly, our argument suggests that at exponentially larger values of $r$, the original conclusion of \cite{Adams:2012yi} is correct after all.


In order to further verify the agreement between holographic and the scaling arguments of the previous section, we solve the gauge field equation of motion to first order in $D$. The behavior of $\mathcal{A}_t$ near the boundary looks like
\begin{equation}
    \mathcal{A}_t  \sim \alpha(k)+\mathcal{O}(r)+r^{1-\frac{3D}{32}}(\beta(k)+\mathcal{O}(r))
\end{equation}
This tells us that the dimension of the dual operator scales like $[\mathcal{J}_t] = [r^{1-\frac{3D}{32}}\mathrm{d}t] = 2$. This is in agreement with the non-renormalization theorem and as a consequence (\ref{eq:ZJJ}).

\subsection{Finite chemical potential, magnetic field and temperature}
One of the advantages to holography is that it allows us to turn on additional background fields or sources that introduce more explicit scales into the problem.   This can be challenging when trying to organize conventional RG, but is no more difficult to analyze than what we did above.  As a concrete example, we will study how the weak marginal disorder affects the system when it is at a finite chemical potential $\mu_0\ll \Lambda$. Now the background clean geometry is no longer $\mathrm{AdS}_4$ but rather Reissner-Nordstrom-AdS (AdS-RN) with a charged black hole \cite{Hartnoll:2016apf}.  The gauge fixed background metric and gauge field solutions are \cite{Hartnoll:2016apf}
\begin{subequations}
\begin{align}
    \mathrm{d}s^2 &= -\frac{f(r)}{r^2}\mathrm{d}t^2+\frac{\mathrm{d}r^2}{f(r)r^2}+\frac{\mathrm{d}\vec{x}^2}{r^2}\\
    \mathcal{A}_t(r) &= \mu_0\bigg(1-\frac{r}{R}\bigg)
\end{align}
\end{subequations}
where 
\begin{equation}
    f(r) = 1-\bigg(1+\frac{R^2\mu^2}{4}\bigg)\bigg(\frac{r}{R}\bigg)^3+\frac{R^2\mu^2_0}{4}\bigg(\frac{r}{R}\bigg)^4
\end{equation} is the emblackening factor with the location of horizon given by $R = \frac{2\sqrt{3}}{\mu_0}$.

The near-horizon regime of this geometry is of the most interest to us as it controls the IR behaviour of the theory. For this regime, we switch from $r$ to a new coordinate
\begin{equation}
    \zeta = \frac{1}{R-r}\frac{R^2}{6}.
\end{equation}
Under this transformation and $x\xrightarrow[]{}c_0 x$ we find that the near-horizon geometry becomes $\mathrm{AdS}_2\times \mathbb{R}^2$ with the gauge field solution given by 
\begin{subequations}
\begin{align}
    \mathrm{d}s^2 &= -\frac{\mathrm{d}t^2}{6\zeta^2}+\frac{\mathrm{d}\zeta^2}{6\zeta^2}+\frac{c_0}{R^2}\mathrm{d}\vec{x}^2\\
    \mathcal{A}_t(\zeta) &= \frac{\mu_0R}{6\zeta}\label{5.12b}
\end{align}
\end{subequations}
In these coordinates, the horizon is at $\zeta\xrightarrow[]{}\infty$. Now adding the disorder implies perturbing the background gauge field solution (\ref{5.12b}) at order $O(D)$. We find that at first order in $D$,
\begin{equation}
    \mathcal{A}_t(k,\zeta) \sim \sqrt{2\pi}(\zeta\mu_0)^{-\frac{\mu_0+\sqrt{8k^2+\mu^2_0}}{2\mu_0}}
\end{equation}
where the overall $O(1)$ normalization constant is unimportant. Using this we arrive at the homogeneous correction at order $O(D)$ to the metric:
\begin{equation}
    \mathrm{d}s^2 = -\frac{1+Df(\zeta)}{6\zeta^2}\mathrm{d}t^2+\frac{1+Dh(\zeta)}{6\zeta^2}\mathrm{d}\zeta^2+c_0\frac{1+Dg(\zeta)}{R^2}\mathrm{d}\vec{x}^2
\end{equation}
where
\begin{subequations}
\begin{align}
    f(\zeta) &= \frac{9c_0\zeta\mu(1+\log(\zeta\mu))-16c_1\log(\zeta\mu)^2+9c_0\log(\zeta\mu)^2\text{li}(\zeta\mu)}{48\zeta\mu \log(\zeta\mu)^2}\\
    h(\zeta) &= -\frac{9c_0\zeta\mu(2+5\log(\zeta\mu))+80c_1\log(\zeta\mu)^2-45c_0\log(\zeta\mu)^2\text{li}(\zeta\mu)}{48r\mu \log(\zeta\mu)^2}\\
    g(\zeta) &= -\frac{c_1}{\zeta\mu}+c_2-\frac{c_0}{32\log(\zeta\mu)^2}-\frac{3c_0}{8\log(\zeta\mu)}+\frac{9c_0\text{li}(\zeta\mu)}{16\zeta\mu}
\end{align}
\end{subequations}
where $\text{li}(x)$ is the logarithmic integral function whose asymptotic behavior is $\lim_{x\xrightarrow{}\infty}\mathrm{li}(x) = \frac{x}{\log(x)}$ and $c_1$ and $c_2$ are constants of integration. Note that all the perturbations are finite and we have no divergence as $\zeta\xrightarrow[]{}\infty$. This tells us that unlike the $\mu_0 = 0$ case, the disorder becomes \emph{strictly} irrelevant \cite{PhysRevLett.108.241601}. In order to fix $c_0$, $c_1$ and $c_2$ we assume the metric smoothly deforms from $\mathrm{AdS}_4$ to $\mathrm{AdS}_2\times \mathbb{R}^2$ over a length scale of $\delta r \sim \frac{O(1)}{R}$ and we match the metric coefficients. 
\begin{equation}
    \mathrm{d}s^2 = -\frac{1}{6\zeta^2}\mathrm{d}t^2+\frac{1}{6\zeta^2}\mathrm{d}\zeta^2+c_0\frac{1+Dc_2}{R^2}\mathrm{d}\vec{x}^2
\end{equation}
where 
\begin{subequations}
\begin{align}
    c_0 &\sim \frac{1}{4}\log \bigg(\frac{\Lambda}{\mu_0}\bigg)\log \bigg(\log \bigg(\frac{\Lambda}{\mu_0}\bigg)\bigg)\\
    c_2 &\sim \frac{3}{32}\log \bigg(\frac{\Lambda}{\mu_0}\bigg)-\frac{3}{80}
\end{align}
\end{subequations}
The above result of the disorder becoming \emph{strictly irrelevant} (not just marginally irrelevant) at finite $\mu_0$ also holds if we had instead put the system at finite magnetic field $B \sim O(1)$ or at a finite temperature $T$, so long as $TR \ll 1$ (which ensures the $\mathrm{AdS}_2\times\mathbb{R}^2$ geometry at intermediate scales).  The fact that the coefficient of $\mathrm{d}\vec{x}^2$ has increased means that disorder has perturbatively increased the entropy density of the low temperature field theory in the IR.

We also briefly relate our results to some older work in the literature. In \cite{Hartnoll:2014gaa}, it was proposed that $\mathrm{AdS}_2\times \mathbb{R}^2$ geometries would admit inhomogeneous deformations; see \cite{Donos:2014yya} for follow-up numerics which argue instead that $\mathrm{AdS}_2\times\mathbb{R}^2$ is stable.  Our results appear consistent with the latter conclusion, though we are not certain whether our approach is sufficiently high order in nonlinear terms to recover the conjectured effect \cite{Hartnoll:2014gaa}.  Interestingly enough, it appears to be the zero density systems (which seemed not to admit highly inhomogeneous IR geometries \cite{Chesler:2013qla,OKeeffe:2015qma}) that lead to highly disordered IR geometries. It would be interesting to test this prediction in future numerics.


\section{Scalar disorder}\label{sec:scalar}
Having confirmed the agreement between our heuristic field theoretic arguments and the minimal holographic theory with charge disorder, we now move on to the theory with scalar disorder which was studied in \cite{Hartnoll:2014cua,Ganesan:2020wzm}.  
We start by introducing the original model. Suppose that this CFT has an operator $\mathcal{O}$ of dimension 
\begin{equation}
   [ \mathcal{O}] = \Delta = \frac{d}{2}+1.
\end{equation}
Here, we will assume that $\mathcal{O}$ is spin zero. For a spin-0 (scalar) operator of this special dimension,
\begin{equation}
 \langle \mathcal{O}(x,t)\mathcal{O}(0,0)\rangle_{\mathrm{CFT}} = \frac{C_{\mathcal{OO}}}{(x^2+t^2)^{1+d/2}}   \label{eq:COO}
\end{equation} 
Now again consider perturbing the CFT Lagrangian density by a space-dependent but time-independent coupling $h(x)$ 
\begin{equation}
    \mathcal{L} \rightarrow \mathcal{L}+ h(x)\mathcal{O}(x,t). \label{eq:lagrangian}
\end{equation}
Note that $[h]=d/2$.  We similarly take the disorder to be drawn from a Gaussian distribution
\begin{equation}
    \overline{h(x)h(x')} = D \times G(\Lambda|x-x'|). \label{eq:variance}
\end{equation}
The replicated action in this case resembles (\ref{eq:seff}) except with the operator $\mathcal{J}_t$ replaced by $\mathcal{O}$. The relevant term in the OPE now takes the form:
\begin{equation}\label{eq:OOPE}
    \mathcal{O}^a(x_1,t_1)\mathcal{O}^b(x_2,t_2) \supset \frac{\delta_{ab}C_{\mathcal{OO}T}}{C_{TT}} \frac{t_{12}^2T_{tt}(x_1,t_1)}{(x_{12}^2+t_{12}^2)^{\frac{3}{2}}}  + \cdots 
\end{equation}
Using this we see that the expression for $Z_t$ (\ref{eq:Zt}) remains the same as well with $C_{\mathcal{JJ}T}$ replaced by $C_{\mathcal{OO}T}$ in agreement with \cite{Hartnoll:2014cua, Aharony:2018mjm}. Note that in any CFT, \cite{Aharony:2018mjm} \begin{equation}
    C_{\mathcal{OO}T} < 0,
\end{equation}
and so $z\ge 1$, consistent with all the understood quantum field theories that we are aware of. 

\subsection{Heuristic argument and comparison to holography}
In general, it is far from clear that the OPE channel $\mathcal{OO}T$ is the only contribution to the RG flow of $D$. However, there is one class of CFTs for which we can assure this is the case: a putative CFT described by gauge-gravity duality \cite{Hartnoll:2016apf}.  In a nutshell, the holographic correspondence states that the CFT (if it is of a suitable large $N$ matrix type) is dual to a semiclassical gravity theory in asymptotically anti-de Sitter (AdS) space in one higher dimension.  The scalar operator $\mathcal{O}$ is dual in the bulk to a scalar field $\phi$, and the stress tensor $T_{\mu\nu}$ is dual to the spacetime metric $g_{\mu\nu}$.  The holographic action controls CFT data, such as scaling dimensions and OPE coefficients.  The bulk holographic model \begin{equation}
    S_{\mathrm{bulk}} = \int \mathrm{d}^{d+2}x_{\mathrm{bulk}} \left(R + \frac{d(d+1)}{L^2} - \frac{1}{2}(\nabla \phi)^2 - \frac{m^2}{2}\phi^2  \right) \label{eq:bulk}
\end{equation}
which was studied in \cite{Hartnoll:2014cua} corresponds to a CFT where the only third order OPE coefficients (of relevance) are $TTT$ and $\mathcal{OO}T$; these coefficients can be read off from the action by computing (schematically) $\delta g^3$ and $\delta g\delta \phi^2$ terms in (\ref{eq:bulk}) upon expanding around AdS spacetime (a saddle point of this action).  By suitably tuning $m^2$, we can arrange for Harris-marginal disorder.  

A quick argument for the fact that $D$ can become marginally irrelevant, which we repeat from \cite{Ganesan:2020wzm}, is instructive.  
The Harris criterion \cite{Harris_1974} for the marginality of scalar disorder in $d+1$ dimensions with dynamical critical exponent $z$ is given by 
\begin{equation}
    \Delta = \frac{d}{2}+z. \label{eq:harris}
\end{equation}
In our scalar theory, we will have $\Delta= \frac{d}{2}+1$ and $z=1$ at the UV fixed point; from (\ref{eq:lifshitz}), we see that $z$ will slightly change. Unless $\Delta$ happens to be rescaled by the exact same amount (which would seem to require one parameter's worth of fine tuning), then we should anticipate that at order $D$, (\ref{eq:harris}) will no longer hold, and so disorder will be either marginally irrelevant or relevant.

In the holographic computation of \cite{Ganesan:2020wzm}, it was found that $\Delta$ increased by a larger amount at order $D$ than $z$, and hence the scalar disorder was marginally irrelevant. This can be seen from the relation between the mass of the bulk scalar field $m$ and the dimension of the dual operator $\Delta$:
\begin{equation}
    \Delta(\Delta-d-z) = m^2
\end{equation}
$m^2=-\frac{d}{2}(\frac{d}{2}+1)$ is fixed by requiring the operator be Harris-marginal to begin with. Using this we find that for $z\approx 1$:
\begin{equation}\label{eq:hol_ano_scalar}
    \Delta \approx \bigg(\frac{d}{2}+1\bigg)+\bigg(\frac{d}{2}+1\bigg)(z-1)+\mathrm{O}((z-1)^2)
\end{equation}
This tells us that the disorder becomes marginally irrelevant.

Therefore with the normalization of $\phi$ above, and in $d=1$, one finds that for energy scales just below $\Lambda$, \begin{equation}
    z\approx 1+\frac{D}{8} = 1 + \frac{D|C_{\mathcal{OO}T}|}{C_{TT}}.
\end{equation}
However, our argument shows that this value of $z$ flows, because $D$ flows. 

In the holographic calculation of \cite{Ganesan:2020wzm}, determination of these flowing quantities relied on a subtle heuristic argument. Disorder arises due to random boundary conditions on the bulk field $\phi$; one must (approximately) construct the back-reacted geometry in the presence of disorder and solve the Einstein equations all the way to a zero temperature horizon in the infrared.  So one must account for non-perturbative effects in $1/D$, which only arise at the energy scale \begin{equation}
    \Lambda_* \sim \Lambda \exp\left[-\frac{C_{TT}}{D|C_{\mathcal{OO}T}|}\right],
\end{equation}
 a scale which was not accessible in the resummed perturbative argument of \cite{Hartnoll:2014cua}. From our RG perspective, this non-perturbative effect is a conventional, one-loop phenomenon of marginally irrelevant disorder.

\subsection{Correction to two-point function}\label{COOcorr}
The calculation for the scalar case proceeds similar to the vector calculation.  We begin with the rescaled correlation function:
\begin{equation}\label{eq:resOO}
    \langle \mathcal{O}(x,t)\mathcal{O}(0,0)\rangle_{\mathrm{CFT}} = \frac{C_{\mathcal{OO}}Z_{\mathcal{OO}}}{(x^2+(Z_tt)^2)^{1+d/2}}
\end{equation}
Our goal is to determine $Z_{\mathcal{OO}}$; we have already determined $Z_t$ in the earlier discussion above.  In this spirit we look at the following two-point function:
\begin{align}\label{eq:414}
    \langle \mathcal{O}^a(0,0)\mathcal{O}^a(x,t)\rangle_D = \langle \mathcal{O}(0,0)\mathcal{O}(x,t)\rangle_{\mathrm{CFT}} + \frac{D}{2}\int \mathrm{d}^dx_1\mathrm{d}t_1\mathrm{d}t_2\langle \mathcal{O}(0,0)\mathcal{O}(x,t)\mathcal{O}(x_1,t_1)\mathcal{O}(x_1,t_2) \rangle_{\mathrm{CFT}}+\cdots 
\end{align}
(no sum over $a$) where higher order terms are not of interest to us. Now plugging in the OPE given in (\ref{eq:OOPE}), we get
\begin{align}
    \langle \mathcal{O}^a(0,0)\mathcal{O}^a(x,t)\rangle_D = &\langle \mathcal{O}(0,0)\mathcal{O}(x,t)\rangle_{\mathrm{CFT}} + \frac{DC_{\mathcal{OO}T}\log  b}{C_{TT}}\int \mathrm{d}^dx_1\mathrm{d}t_1\langle \mathcal{O}(0,0)\mathcal{O}(x,t)T_{tt}(x_1,t_1)\rangle_{\mathrm{CFT}}
\end{align}
where the three point function is given by 
\begin{equation}\label{eq:OOT}
    \langle O(x_1)O(x_2)T_{\mu\nu}(x_3)\rangle = C_{OOT}\frac{V_{\mu}V_{\nu}-\frac{\delta_{\mu\nu}}{d}V_{\alpha}V_{\alpha}}{x_{13}^{d-1}x_{23}^{d-1}x_{12}^{2\Delta-d+1}}
\end{equation}
with $V^{\alpha} = \frac{x^{\alpha}_{13}}{x^2_{13}}-\frac{x^{\alpha}_{23}}{x^2_{23}}$. 
We now use the result given in \cite{Aharony:2018mjm} to evaluate the integral over the three-point function to get,
\begin{equation} \label{eq:417}
  \int \mathrm{d}^dx_3 \mathrm{d}t_3  \left\langle \mathcal{O}(0,0)\mathcal{O}(x_2,t) T_{tt}(x_3,t_3)\right\rangle_{\mathrm{CFT}} = C_{OO}\frac{d+2}{d+1}\frac{x^2+t^2-(d+1)t^2}{(x^2+t^2)^{\frac{d}{2}+2}}
\end{equation}
However, it is important to note that (\ref{eq:OOT}) does not include contact terms. These terms go as \cite{Srednicki:2007qs, Peskin:1995ev}: 
\begin{equation}\label{eq:contact}
    \langle T_{tt}(x)\mathcal{O}(x_1)\mathcal{O}(x_2)\rangle \supset \mathcal{C}(\delta^{(d+1)}(x-x_1)+\delta^{(d+1)}(x-x_2))\langle \mathcal{O}(x_1)\mathcal{O}(x_2)\rangle
\end{equation}
We fix the coefficient of the contact term by requiring that the equal $x$ correlator does not gain any logarithmic corrections. This can be motivated by looking at the correction to the action in (\ref{eq:seff}) containing two time integrals leading to no `$\log t$' dependence for disorder profiles with short range correlations (disorder at different points are not necessarily statistically independent), evaluating the integral in (\ref{eq:414}) we obtain
\begin{align}
    \int \mathrm{d}^dx_1\mathrm{d}^dx_2 \mathrm{d}t_1\mathrm{d}t_2 \;G(\Lambda|x_1-x_2|)\langle \mathcal{O}(0,0)\mathcal{O}(x_1,t_1)\rangle \langle \mathcal{O}(0,t)\mathcal{O}(x_2,t_2)\rangle &\sim \Lambda_{\textrm{CFT}}^{d+2},\\
    \int \mathrm{d}^dx_1\mathrm{d}^dx_2 \mathrm{d}t_1\mathrm{d}t_2 \;G(\Lambda|x_1-x_2|)\langle \mathcal{O}(0,0)\mathcal{O}(x_1,t_1)\rangle \langle \mathcal{O}(x,0)\mathcal{O}(x_2,t_2)\rangle &\sim \frac{\Lambda_{\textrm{CFT}}}{x^{d+1}}+\frac{\log(x\Lambda)}{x^{d+2}}+\cdots.
\end{align}
The former integral indicates that without further OPE corrections in the scalar channel (which we do not consider in this paper), the correction to the scaling exponent $\Delta$ is fixed by the requirement that that the equal-$x$ correlator does not gain an anomalous dimension with $t$.

Combining (\ref{eq:417}) with (\ref{eq:contact}), we find that this happens when (recall $\Delta=1+d/2$):
\begin{equation}
    \mathcal{C} = \frac{d\Delta}{d+1}.
\end{equation}
Hence we find
\begin{align}
    \langle \mathcal{O}^a(0,0)\mathcal{O}^a(x,t)\rangle_D = \frac{C_{\mathcal{OO}}}{(x^2+t^2)^{1+\frac{d}{2}}}\bigg(1+\frac{DC_{\mathcal{OO}T} (d+2)x^2\log b}{C_{TT}(x^2+t^2)}\bigg)
\end{align}
We find that with the contact term (\ref{eq:contact}), the results for equal $x$ and $t$ correlators are in line with holographic results from \cite{Ker_nen_2017}. For the vector case, we did not need to worry about contact terms since $\mathcal{J}$ is a conserved current even under renormalization.

Now plugging in the rescaled correlator (\ref{eq:resOO}), we can see that the divergence appearing in the renormalized correlator is removed by choosing
\begin{equation}
    Z_{\mathcal{OO}} = 1-\frac{(d+2)DC_{OOT}\log  b}{C_{TT}}
\end{equation}
This leads to the scalar operator gaining an anomalous dimension given by 
\begin{equation}\label{eq:ano_scalar}
    \gamma_{\mathcal{O}} = \frac{1}{2}\frac{\partial Z_{\mathcal{OO}}}{\partial \log b} = -\frac{(d+2)D C_{\mathcal{OOT}}}{2C_{TT}}
\end{equation}
The final step is to find the relation between $D^{\text{bare}}$ and $D$ (renormalized):
\begin{align}
    D^{\text{bare}} \approx D^R\bigg(1-\frac{dD^RC_{\mathcal{OO}T}}{C_{TT}}\log b\bigg)
\end{align}
Or, we can re-write it as
\begin{align}
    D^R \approx D^{\text{bare}}\bigg(1+\frac{dD^{\text{bare}}C_{\mathcal{OO}T}}{C_{TT}}\log b\bigg)
\end{align}
Let us put this in the form of a conventional Wilsonian RG equation. 
Assuming that the only important contribution to the flow of $D$ arises due to the stress tensor channel, we conclude that 
\begin{equation}
    b\frac{\partial D}{\partial b} = -\frac{dD^2|C_{\mathcal{OO}T}|}{C_{TT}}
\end{equation}
We have used that $C_{\mathcal{OO}T} < 0$. At energy scale $E$, we conclude that the effective disorder strength is given by \begin{equation}
    D(E) = \dfrac{D(\Lambda)}{\displaystyle 1 + D(\Lambda) \frac{d|C_{\mathcal{OO}T}|}{C_{TT}}\log b} \label{eq:Dflow}
\end{equation}
Hence, disorder is marginally irrelevant.  We can also see what happens to the flow of the dynamical critical exponent $z$.  Upon integrating (\ref{eq:dlogt}) using (\ref{eq:Dflow}), we find that 
\begin{equation}
   -\frac{\mathrm{d}\log t}{\mathrm{d}\log b} = 1 + \dfrac{\displaystyle D(\Lambda)\frac{|C_{\mathcal{OO}T}|}{C_{TT}}}{\displaystyle 1 + D(\Lambda) \frac{d|C_{\mathcal{OO}T}|}{C_{TT}}\log b}
\end{equation}
leading to
\begin{equation}
    -\log t = \log b + \frac{1}{d}\log \left( 1 + D(\Lambda) \frac{d|C_{\mathcal{OO}T}|}{C_{TT}}\log b\right)
\end{equation}
The effective dynamical critical exponent at energy scale $E<\Lambda$ is given by \begin{equation}
    z(E) = -\frac{\log t}{\log b} = 1 + \dfrac{\displaystyle \log \left( 1 + D(\Lambda) \frac{d|C_{\mathcal{OO}T}|}{C_{TT}}\log \frac{\Lambda}{E}\right)}{\displaystyle d\log \frac{\Lambda}{E}} \label{eq:zflow}
\end{equation}

The true infrared theory has irrelevant disorder given by (\ref{eq:Dflow}), and the unusual effective dynamical critical exponent (\ref{eq:zflow}). These two results are precisely what was found in \cite{Ganesan:2020wzm}. We also see that the anomalous dimension is in agreement between the two theories from \eqref{eq:hol_ano_scalar} and \eqref{eq:ano_scalar} indicating that that what we have in (\ref{eq:contact}) is indeed the correct contact term.

\section{Conclusions}\label{sec:conclusions}
We have discussed, both using gauge-gravity duality and scaling arguments inspired by conformal perturbation theory, the RG flows of Harris-marginal disorder in strongly coupled CFTs.  We find that scalar disorder is marginally irrelevant while charge disorder is marginally relevant, when the dominant coupling is of these operators to the stress tensor.  This assumption is indeed the case in the matrix-large-$N$ field theories that have weakly coupled holographic dual descriptions. 

Our work provides a more physical picture to the more heuristic construction of bulk geometries advocated in \cite{Ganesan:2020wzm}; in particular, in this paper we have been able to precisely match the field theoretic flows of coupling constants to the evolution of the bulk geometry in the 2d model with charge disorder.  As such, we conclude that the speculative Lifshitz fixed points of \cite{Hartnoll:2014cua} do not exist as genuine IR fixed points (at least at weak coupling).  It may, unfortunately, be quite challenging to resolve the endpoint of the RG flow, since as we emphasized above, the energy scales where the ``tree level" calculation of $z$ becomes inaccurate are exponentially small.  In this sense, the Lifshitz fixed point can appear for a wide range of scales.

In more practical physical systems, it is less clear whether our assumption is justified.  In principle, one may wish to perform loop calculations within conformal perturbation theory to try and compute the RG flow of disorder etc. without resorting to general scaling arguments, as we did in this paper.  We leave a more systematic development of that method to future work.

Our approach provides an interesting alternative to the more conventional methods for studying the interplay of disorder and strong interactions. Usually in the field theory literature, one either starts with a theory with a \emph{vector} large-$N$ limit, which is for practical purposes weakly coupled in any spatial dimension, or by doing an $\epsilon$-expansion about an upper critical dimension.  Each of these approaches relies on conventional Feynman diagrammatics and the validity of this approach at a strongly coupled fixed point is not clear.   Our approach does not assume anything about the strength of coupling constants, instead assuming the dominant OPE channels.  In this sense it complements existing approaches.  We hope that our approach can provide a useful new tool for future studies of strongly interacting and disordered field theories.

\section*{Acknowledgements}
We thank Sean Hartnoll for useful comments on a draft.  KG and AL were partially supported by a Research Fellowship from the Alfred P. Sloan Foundation under Grant FG-2020-13795, and by the Gordon and Betty Moore Foundation's EPiQS Initiatve under Grant GBMF10279. LR was supported by the Simons Investigator Award from the James
Simons Foundation.




\bibliography{thebib}

\end{document}